\documentclass[aps,
notitlepage,twocolumn,showpacs,superscriptaddress,nofootinbib,preprintnumbers]{revtex4-1}

\usepackage{graphicx}
\usepackage{dcolumn}
\usepackage{bm}
\usepackage{dsfont}
\usepackage{amsmath,amssymb}
\usepackage{braket}

\newcommand{\dd}{\mathrm{d}}

\usepackage[utf8]{inputenc}
\usepackage{float}
\usepackage{amsmath,amssymb,mathtools,bm}
\usepackage{dsfont} 
\usepackage{framed}
\usepackage[bbgreekl]{mathbbol}
\usepackage{slashed}
\usepackage[unicode=true,pdfusetitle,
bookmarks=true,bookmarksnumbered=false,bookmarksopen=false,breaklinks=false,pdfborder={0 0 1},backref=false,colorlinks=true]{hyperref}

\usepackage{tikz,tikz-feynman,contour}
\usepackage{subcaption}

\usepackage{soul}

\definecolor{plotgreen}{rgb}{0.0, 0.65, 0.0}

\begin{document}

\preprint{APS/XX}

\title{Decoherence effects in entangled fermion pairs at colliders}%

\author{Rafael Aoude}%
\email{rafael.aoude@ed.ac.uk}
\affiliation{Higgs Centre for Theoretical Physics,
School of Physics and Astronomy,\\
The University of Edinburgh, Edinburgh EH9 3JZ, Scotland, UK}%

\author{Alan J. Barr}%
\email{alan.barr@physics.ox.ac.uk}
\affiliation{ 
Department of Physics, Keble Road, University of Oxford, OX1 3RH
Merton College, Merton Street, Oxford, OX1 4JD}%

\author{Fabio Maltoni}
\email{fabio.maltoni@unibo.it}
\affiliation{ 
Dipartimento di Fisica e Astronomia, Universit\`a di Bologna, Via Irnerio 46, 40126 Bologna, Italy}%
\affiliation{Centre for Cosmology, Particle Physics and Phenomenology (CP3), Universit\'e Catholique de
Louvain, B-1348 Louvain-la-Neuve, Belgium}%

\author{Leonardo Satrioni}%
\email{leonardo.satrioni@studio.unibo.it}
\affiliation{ 
Dipartimento di Fisica e Astronomia, Universit\`a di Bologna, Via Irnerio 46, 40126 Bologna, Italy}%
\affiliation{Higgs Centre for Theoretical Physics,
School of Physics and Astronomy,\\
The University of Edinburgh, Edinburgh EH9 3JZ, Scotland, UK}%

\date{\today}

\begin{abstract}

Recent measurements at the Large Hadron Collider have observed entanglement in the spins of $t\bar t$ pairs. The effects of radiation, which are expected to lead to quantum decoherence and a reduction of entanglement, are generally neglected in such measurements. In this work we calculate the effects of decoherence from various different types of radiation for a maximally entangled pair of fermions --- a bipartite system of qubits in a Bell state.  
We identify the Kraus operators describing the evolution of the open quantum system with the integrated Altarelli-Parisi splitting functions.
\end{abstract}

\maketitle


\section{Introduction}

Entanglement is a defining feature of quantum mechanics and a central concept in quantum information theory and quantum computing~\cite{Horodecki:2009zz}. 
Recently, it has been realized that spin entanglement between pairs of massive elementary particles produced at colliders can be accessed experimentally through their decays mediated by the chiral weak force~\cite{Afik:2020onf,Severi:2021cnj,Aoude:2022imd,Afik:2022kwm,Afik:2022dgh,Barr:2022wyq,Severi:2022qjy,AshbyPickering:2022umy,Aguilar-Saavedra:2022wam,Aguilar-Saavedra:2022mpg,Fabbrichesi:2022ovb,Aguilar-Saavedra:2022uye,Fabbri:2023ncz,Aoude:2023hxv,Han:2023fci,Fabbrichesi:2023cev,Sakurai:2023nsc,Altomonte:2023mug,Afik:2024uif,Aguilar-Saavedra:2024vpd,Aguilar-Saavedra:2024whi,Grabarczyk:2024wnk,Morales:2024jhj,White:2024nuc,Altomonte:2024upf,Han:2024ugl,Liu:2025qfl}.\!\footnote{A recent review of such proposals to measure quantum entanglement and related quantities at colliders may be found in Ref.~\cite{Barr:2024djo}.} 
Among the simplest and most studied cases are pairs of fermions, such as $\tau$ leptons or top quarks, whose decay products can be fully reconstructed in detectors.  
The first observations of quantum entanglement at the TeV scale have recently been reported by the ATLAS~\cite{ATLAS:2023fsd} and CMS~\cite{CMS:2024pts,CMS:2024zkc} collaborations, in measurements of spin correlations between top quarks and their antipartners.\!\footnote{While spin correlations of particle pairs have been studied at colliders for decades, the recent quantum-information perspective recasts them in terms of the underlying quantum state, enabling the quantification of their quantumness and and providing complementary information to the dynamics of the fundamental interactions and possible physics beyond
the Standard Model. 
Top quarks provide an ideal laboratory for such studies, since they typically decay before hadronizing and thus before QCD interactions can erase their spin correlations~\cite{Falk:1993rf,PhysRevD.43.1500}. 
Indeed, the characteristic decay time $1/\Gamma_t \sim 1/1.42~\mathrm{GeV}^{-1}$ is shorter than both the hadronization timescale $1/\Lambda_{\rm QCD} \sim 1/250~\mathrm{MeV}^{-1}$ and the spin-decorrelation time $1/(\Lambda_{\rm QCD}^2/m_t) \sim 0.36~\mathrm{MeV}^{-1}$~\cite{Mahlon:2010gw}.}

In these first measurements, the $t\bar{t}$ system was modeled as a closed system at the point of the quark decays. However, top quarks may radiate gluons or photons in the short period of time before decaying, 
leading to a reduction in quantum spin information, \textit{i.e.}, \textit{decoherence}. 
It is generally been expected that next-to-leading (NLO) and next-to-next-to-leading (NNLO) order corrections to spin correlations in QCD are small~\cite{Behring:2019iiv,Czakon:2020qbd,Mazzitelli:2021mmm} and therefore have therefore been assumed to have a negligible effect in entanglement measures. The reader is referred to Refs.~\cite{Groote:2008ux,Groote:2010zf,Groote:2009zk,Bernreuther:2015yna,Frederix:2021zsh} for further works on the spin-state of the top quarks and the angular
distributions of the daughter leptons.
Similarly, NLO QED effects in the quantum information studies of $h,Z\to \tau^+\tau^-$ have not been included. With an increasing program of future quantum measurements planned at high-energy colliders it is timely to revisit these assumptions.

Decoherence can be studied by recognizing that realistic quantum systems are always embedded in some environment. This interaction with the system results in `leakage of information' to the environment, decreasing the entanglement between the components of the system under test~\cite{Zeh1970OnTI,Zurek:1981xq,Zurek:1982ii,Paz,zurek2003decoherencetransitionquantumclassical}. 
In particle physics, decoherence has been explored in the context of flavour entanglement for Kaon systems~\cite{Bertlmann:2004yg}, effective field theory~\cite{Burgess:2024heo,Salcedo:2024smn,Salcedo:2024nex} and soft radiation~\cite{Carney:2017jut,Carney:2017oxp,Carney:2018ygh,Neuenfeld:2018fdw,Semenoff:2019dqe}.\footnote{The reader is referred to e.g.~\cite{Schlosshauer:2019ewh} for a more complete overview of decoherence in general.}

In this paper we formalize the effects of decoherence from radiation at high-energy colliders and calculate the size of the expected loss of entanglement. Our approach is general and can be applied to any fermion-antifermion pair in any quantum state. As a case study we consider a fermion-antifermion pair, {\it e.g.}, $t\bar t$ or $\tau^+ \tau^-$,   in a maximally
entangled state, such as that originating from the decay of a scalar state. We treat the $f\bar{f}$ pair as an open quantum system -- a bipartite system of spin qubits -- and the additional radiation is interpreted as an interaction with the unobserved `environment'. We calculate change in the concurrence due to the radiation, allowing for various different coupling strengths and Lorentz structures. Finally, we identify a correspondence between Kraus operators in spin space and Altarelli-Parisi splitting functions.

\section{Quantum Maps for Open Systems}
\label{sec:Maps_Review}
Let us first review some of the properties of quantum maps that will be useful in describing decoherence.\footnote{See e.g.~\cite{Aolita_2015} for a review.} The evolution of a non-closed system can  represented by a map $\mathcal{E}$ which acts on the space of density matrices $\rho$ and satisfies the following properties: \textit{convex linearity, trace preservation, and complete positivity}.
Such maps admit an operator-sum representation,
\begin{equation}
\label{eq:OperatorSumRepresentation}
    \mathcal{E}[\rho] = \sum_j K_j \rho K_j^\dagger, \qquad \sum_j K^\dagger_j K_j =\mathds{1}  ,
\end{equation}
where $K_j$ are known as the Kraus operators. 

In this study we establish and illustrate the connection between quantum maps and radiation in quantum field theory in a system composed of two spin-half particles, which 
can be represented as qubits. 
A single-qubit density matrix can be represented in terms of Pauli matrices as $\rho = \frac{1}{2}\big(\mathds{1} + \bm{a}\cdot \bm{\sigma}\big)$. The corresponding Kraus operators can also be represented in terms of the Pauli and the identity matrices. 
Eq.~\eqref{eq:OperatorSumRepresentation} remains valid 
for a state comprising $N$ qubits where now $j$ runs from 0 to $d_S = 2^N$  and this index can be expressed as a multi-index $r \equiv r_1 \cdots r_N$. 

In some $N$-qubit systems, it is possible to decompose the Kraus operators as a product
\begin{equation}
 K_r = K^{(1)}_{r_1} \otimes \cdots \otimes K^{(N)}_{r_N}
\end{equation}
where each $ K_{r_i}$ acts on the subspace of a single qubit $\mathcal{H}_i$. In this case, we call the map an \textit{independent map}. This means that the system corresponding to each qubit effectively has its own environment, which does not interact with the others. In such cases, we can write the operator-sum representation as
\begin{align}
\mathcal{E}[\rho] 
&
\equiv\mathcal{E}_1\otimes \mathcal{E}_2 \otimes \cdots \otimes \mathcal{E}_N(\rho)\\
&= \sum_{r_1\cdots r_N}  
\big(K^{(1)}_{r_1} \otimes \cdots \otimes K^{(N)}_{r_N}\big)
\rho 
\big(K^{(1)}_{r_1} \otimes \cdots \otimes K^{(N)}_{r_N}\big)^\dagger \nonumber
\end{align}
If this decomposition is not possible, we call it a \textit{collective map}, which corresponds to a process in which at least two of the qubits share the same environment.

For our study, we will focus on bipartite maps but our reasoning is general. We call the two Hilbert subspaces $\mathcal{H}_a$ and $\mathcal{H}_b$, such that 
$\mathcal{H}=\mathcal{H}_a \otimes \mathcal{H}_b$,
with a collective index $r\equiv (r_ar_b)$. We choose Pauli-basis maps and represent the full Kraus operators as
\begin{align}
K_{r_ar_b} \equiv 
\sqrt{{\sf p}_{(r_a,r_b)}}\,
\sigma^{r_a} \otimes \sigma^{r_b}\,,
\end{align}
where the notation clarifies the subspace on which each Pauli matrix acts.  The probabilities ${\sf p}_{(r_a,r_b)}$ must sum to unity.  
The index $r_i$ runs from 0 to 3 where $\sigma^0 = \mathds{1}_2$ and the others are the Pauli matrices $\sigma^i$.
In the specific case in which we can factorize ${\sf p}_{(r_a,r_b)} = {\sf p}_{r_a}\times {\sf p}_{r_b}$, we have an independent map
\begin{align}
K_{r_ar_b} \equiv K_{r_a}\otimes K_{r_b} =(\sqrt{{\sf p}_{r_a}}\,
\sigma^{r_a}) \otimes 
(\sqrt{{\sf p}_{r_b}}\,
\sigma^{r_b}).
\end{align}

When a system is embedded in an environment, the entanglement decreases through system-environment interactions. This is the environmental monitoring and information is being transferred.

We model the interactions with the environment through the emission (and possibly absorption) of soft and collinear radiation which is not \textit{resolvable} by the experimental devices. The literature on soft radiation changing the (momenta) entanglement is extensive~\cite{Carney:2017jut,Carney:2017oxp,Carney:2018ygh,Neuenfeld:2018fdw,Semenoff:2019dqe,Tomaras:2019sjq,Irakleous:2021ggq}, as well as on quantum interference for parton showers~\cite{COLLINS1988794,KNOWLES1988767,Nagy:2007ty,Nagy:2008eq,Nagy:2008ns,Nagy:2012bt,Richardson:2001df,Richardson:2018pvo,Hamilton:2021dyz}. However, decoherence for spinning d.o.f in the recent context collider entanglement has not been studied in detail to the best of our knowledge. As we will see, collinear radiation plays a central role in the decoherence of the fermion spin.

\section{Decoherence in scalar decays to a $f\bar f$ pair }
\label{sec:TopPairDecoherence}

To formalise our approach we choose the simplest possible process and $f\bar f$ state, \textit{i.e.}, the one arising in the decay of a heavy scalar boson and we consider the emission of an arbitrary form of radiation. The decoherence may occur due to soft and/or collinear emission that is unresolved by the detectors. These corrections are reformulated as a quantum map acting on the leading order density matrix, in which the unresolved radiation Hilbert space forms the \textit{environment}.

\begin{figure}[htb!]
    \flushleft
    \tikzfeynmanset{
    every blob={/tikz/fill=blue!30,/tikz/inner sep=2pt,/tikz/minimum size=0.5cm},
    every dot={/tikz/fill=red!30,/tikz/inner sep=2pt}}
    \begin{tikzpicture}
    \begin{feynman}
        \vertex at (0,0) (a) {${\varphi(p)}$};
        \vertex at (1.75,0) (b);
        \vertex at (3.0,1.4) (u) {${f(p_1)}$};
        \vertex at (3.0,-1.4) (d) {${\bar{f}(p_2)}$};
        \diagram*{
            (a) -- [scalar] (b),
            (b) -- [fermion] (u),
            (b) -- [anti fermion] (d),
        };
        \vertex[blob] at (1.75,0) {};
    \end{feynman}%
    \hspace{4cm}
     \begin{feynman}
        \vertex at (0,0) (a) {${\varphi(p)}$};
        \vertex at (1.75,0) (b);
        \vertex at (3.0,-0.2) (c) {${p_k}$};
        \vertex at (3.0,1.4) (u) {${f(p_1)}$};
        \vertex at (3.0,-1.4) (d) {${\bar{f}(p_2)}$};
        \vertex[dot,label={above left:\color{red!50}$g_\Gamma  \Gamma^\mu$}] at (2.25,0.58) (u2) {};
        \diagram*{
            (a) -- [scalar] (b),
            (b) -- [fermion] (u2) --[fermion] (u),
            (b) -- [anti fermion] (d),
            (u2) -- [photon] (c),
        };
    \end{feynman}
    \end{tikzpicture}
    \caption{Scalar boson decay to a $f \bar f$ pair: left diagram represents both tree and one-loop virtual correction level and right diagram real emission (from the fermion and the anti-fermion)}
    \label{fig:HiggsDecayDiagram}
\end{figure}
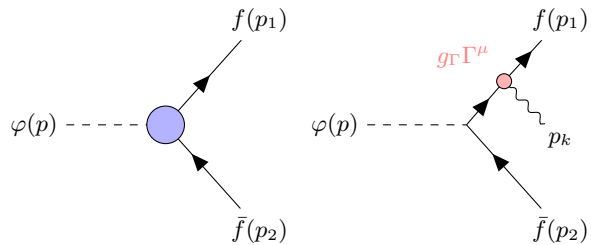%

The tree-level scalar boson decays via the Yukawa interaction  $y_f$ included in the blue blob of Fig.~\ref{fig:HiggsDecayDiagram} with the following $R$-matrix
\begin{align}
R_{\rm LO} =  \frac{4 N_C y_f^2 m_f^2\beta^2}{1-\beta^2}
\begin{pmatrix}
0 &0 &0 &0\\
0 &1 &1 &0\\
0 &1 &1 &0\\
0 &0 &0 &0
\end{pmatrix}\,,
\end{align}
where $N_C$ is the number of colors of the fermions.
In the $f\bar{f}$ production, the trace of this matrix is proportional to the cross-section
while here it is proportional to the decay rate via
$\text{tr}[R_{\rm LO}] \beta/(16\pi m_\varphi) =\Gamma^f_{\rm LO}$.
This density matrix is a function of the available centre-of-mass energy via the velocity $\beta = \sqrt{1-4m_f^2/m_\varphi^2}$ of the fermion  in the scalar boson rest frame and $m_\varphi$ is its mass.
Coming from a singlet decay, the spin density matrix,{\it i.e.}, the normalized $R$-matrix, is that for the 
triplet Bell state.\footnote{In the case of the decay of a pseudo-scalar state, the fermion-antifermion pair is in a spin singlet Bell state, with no change in the following discussion}
\begin{align}
\rho_{\rm LO} = \frac{1}{\text{tr}[R_{\rm LO}]} R_{\rm LO} = |\Psi^+\rangle \langle \Psi^+|\,,
\end{align}
which is a maximally entangled, \textit{i.e.}, the \textit{concurrence} $\mathcal{C}[\rho_{\rm LO}]=1$.
At the next-to-leading order, we consider the real and virtual emission of both scalars and vectors with arbitrary couplings, \textit{i.e.}, scalar/pseudoscalar, vector/axial-vector. The general $g_\Gamma \Gamma^\mu$ interaction represents
\begin{align}
g_{\Gamma}\Gamma^\mu = \{g_S \mathds{1}, g_P \gamma^5, g_V \gamma^\mu, g_A \gamma^\mu\gamma^5\}\,. 
\end{align}
The virtual corrections lead to the same final state Hilbert space while the real emission leads to the extra Hilbert space of the environment. In the scalar and pseudoscalar couplings, we only have extra momentum while the vector and axial include the extra spin degrees of freedom (d.o.f). To obtain the reduced density matrix of $t\bar{t}$, we need to trace\,
over the emitted radiation d.o.f, 
\textit{i.e.}, $\text{tr}_{\mathcal{H}_k}[\cdot ]=\int_{p_k}
\sum_{\sigma=\pm} \langle p_k,\sigma|\cdot |p_k,\sigma\rangle$. 
The reduced density matrix can be written as
\begin{align}
    \rho_{\rm LO+NLO}^{\rm red}
    = {\sf p}_{\rm LO} \, \mathds{1}\rho_{\rm LO}\mathds{1} +
    \bar{\mathcal{E}}_{\rm V} [\rho_{\rm LO}]
    +  \bar{\mathcal{E}}_{\rm R} [\rho_{\rm LO}],
\end{align}
where ${\sf p}_{\rm LO}$ is the Kraus coefficient related to the identity at LO, which is one of the Kraus operators in a full map. The bar on top of map symbol indicates that this constitutes part of the full map, where the probabilities add up to one.  The helicity sum and momentum integration are performed inside the quantum map. The   $\bar{\mathcal{E}}_{\rm V}[\rho]$ and $\bar{\mathcal{E}}_{\rm R}[\rho]$ are the contributions from  virtual (V) and real (R) emission, which we interpret as the following quantum maps
\begin{align}
\bar{\mathcal{E}}_{\rm V} [\rho_{\rm LO}] = {\sf p}_{\rm V} \mathds{1} \rho_{\rm LO} \mathds{1},
\quad
\bar{\mathcal{E}}_{\rm R} [\rho_{\rm LO}] = \sum_j K_{j} \rho_{\rm LO} K^\dagger_{j}.
\end{align}
The virtual contribution in a scalar boson decay is special as it does not change the structure of $\rho_{\rm LO}$. This is because the tree-level amplitude has the simple structure $[\bar{u}(p_1)v(p_2)]$ and the loops can be simplified, using Passarino Veltamn reduction, to the same structure with an overall momenta dependence (See supplementary material for a detailed discussion, which includes~\cite{Passarino:1978jh,Maltoni:2024wyh}). The virtual correction map for massive vector decay (or a $2\to 2$ scattering) is, on the other hand, a non-trivial Kraus map due to the appearance of new structures, such as  a finite dipole term. As the aim of this work is to introduce and illustrate the effects we leave the study of other initial states to future work.
The virtual contribution $p_{\rm V}$ in general display UV and IR divergences. The former is treated in the usual way by a renormalization procedure that does not affect the quantum state, while the IR divergences are canceled by an identity operator from the real emission, as dictated by the KLN theorem~\cite{Kinoshita,Lee:1964is}.

The real emission is different. Let us first split it into two parts where the emission is soft or hard (collinear) with respect to a reference energy scale. The two regions lead to unresolved radiation.
In the former, one can use soft theorems to show that this does not change the spin structure for scalar and vector emissions but do for pseudovector and axial.\footnote{This is true at LO in the soft expansion. Higher-order soft terms are known to contain a spin dependence that will affect the soft part of the map.} 
The hard (collinear) emission does change the spin structure leading to a non-trivial Kraus operator
\begin{align}
\label{eq:Map_Hard_Real}
\bar{\mathcal{E}}_{\rm R} [\rho_{\rm LO}] = 
\bar{\mathcal{E}}^{\rm soft}_{\rm R} [\rho_{\rm LO}] + 
\bar{\mathcal{E}}^{\rm hard}_{\rm R} [\rho_{\rm LO}]\,.
\end{align}
For scalar and vector emission, we have a map with the same structure as the virtual case, while for pseudoscalar and axial cases, 
the $\gamma^5$ leads to an additional Kraus operator ($\mathds{1}\otimes \sigma_3$ or $\sigma_3 \otimes \mathds{1}$)
for this emission and a parameter ${\sf q}^{\rm soft}_5$ which is IR finite.
\begin{align}
\label{eq:Map_Soft_Real}
\bar{\mathcal{E}}^{\rm soft}_{\rm R} [\rho_{\rm LO}] = 
\overbrace{
\underbrace{{\sf p}_{\rm R}^{\rm soft} \mathds{1} \rho_{\rm LO} \mathds{1}}_{\rm  scalar, vector}
+ {\sf q}^{\rm soft}_5\sum_{j\neq \text{id}} K_{j} \rho_{\rm LO} K^\dagger_{j}}^{\rm pseudoscalar, axial}.
\end{align}
These identity coefficients ${\sf p}_{\rm R}^{\rm soft}$ combine to cancel infrared divergences in the same way as occurs in the decay rate. 
The hard emission will have two map structures
\begin{align}
\label{eq:Map_Hard_Real}
\bar{\mathcal{E}}^{\rm hard}_{\rm R} [\rho_{\rm LO}] =
{\sf p}_{\rm R}^{\rm hard}\mathds{1} \rho_{\rm LO} \mathds{1}
+
{\sf q}^{\rm hard}\sum_{j\neq \text{id}} K_{j} \rho_{\rm LO} K^\dagger_{j}\,,
\end{align}
where the structure of the logs in ${\sf p}_{\rm R}^{\rm hard} + {\sf p}_{\rm R}^{\rm soft}$ 
matches well established results in the literature~\cite{Braaten:1980yq}. 
The latter summation of the Kraus operators, which comes with an overall probability ${\sf q}^{\rm hard}$, does not include the identity $\mathds{1}_2\otimes \mathds{1}_2$. Radiation is considered unresolvable if either soft or collinear. Given that the main contribution of real emission comes from the collinear limit, we take this approximation when computing the hard part. Combining all terms, we obtain
\begin{align}
\mathcal{E}_{\rm full} [\rho_{\rm LO}] =
{\sf p}_{\rm id}\,\mathds{1} \rho_{\rm LO} \mathds{1}
+{\sf q}\sum_{j\neq \text{id}} K_{j} \rho_{\rm LO} K^\dagger_{j}\,,
\end{align}
where ${\sf p}_{\rm id} = \left({\sf p}_{\rm LO}+ {\sf p}_{\rm V} + {\sf p}_{\rm R}^{\rm soft} + {\sf p}_{\rm R}^{\rm hard}\right)$ and ${\sf q} = ({\sf q}^{\rm hard} + {\sf q}^{\rm soft}_5)$. Here we have the full map in which the probabilities sum up to one.
In ${\sf p}_{\rm id}$ we have the cancellation of the IR divergences as in NLO corrections to the scalar boson decay~\cite{DREES1990455,Janot:1989jf,Kniehl:1991ze,Braaten:1980yq}. The list of explicit Kraus operators for each case is listed in the appendix. 

\begin{figure}[!htb]
\centering
\includegraphics[width=0.485\textwidth]{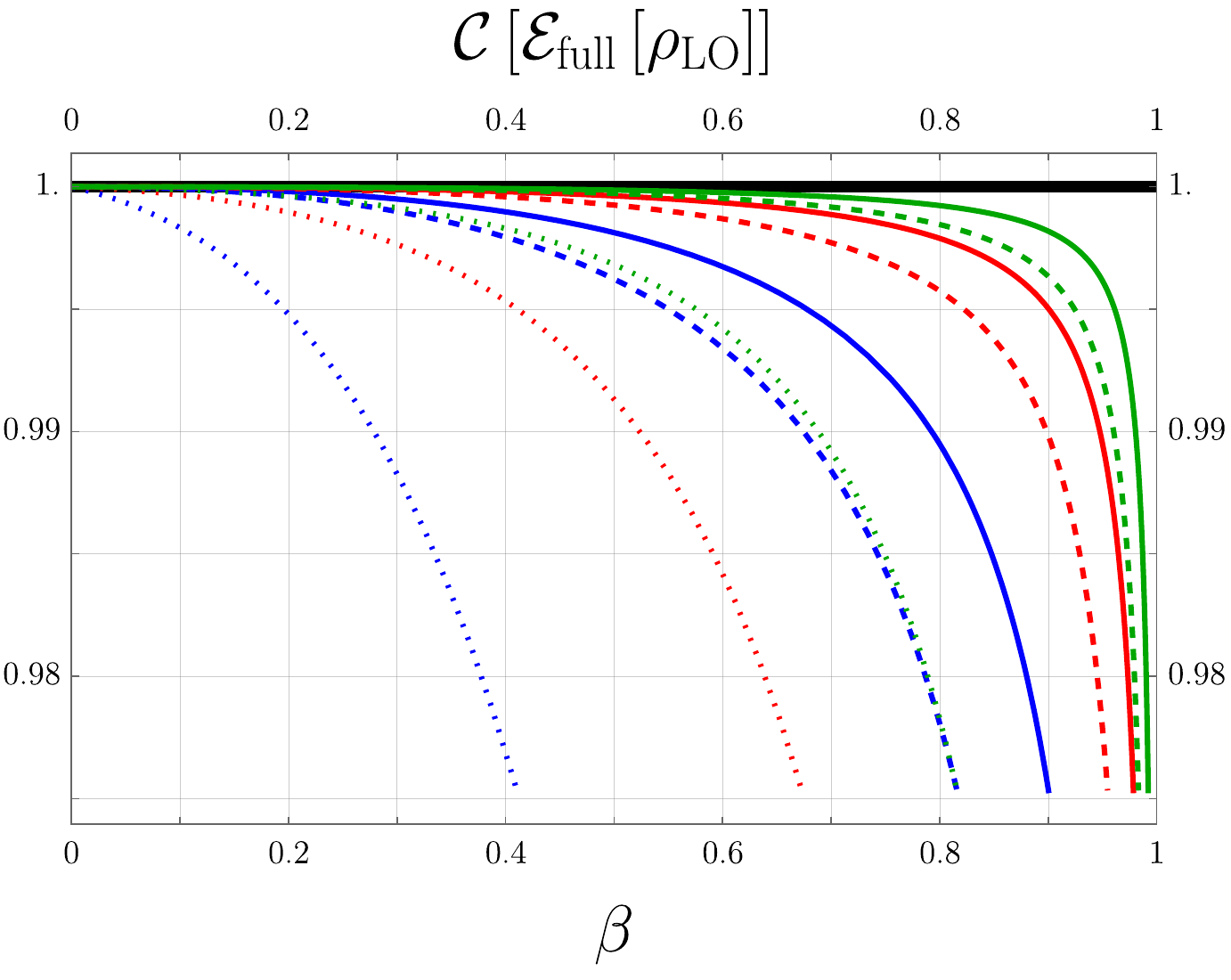}
\caption{Concurrence for a scalar decay to $t \bar t$ due to a collinear emission. The LO and the scalar lines overlap at $\mathcal{C}=1$. Line styles: LO and scalar (thick), pseudoscalar (solid), vector (dashed), axial (dotted) and the different interaction strengths are $\alpha_\Gamma = 1/10$ {\color{blue}(blue)}, $\alpha_\Gamma = 1/50$ {\color{red} (red)} and $\alpha_\Gamma = 1/137$ {\color{plotgreen}(green)}.}
\label{fig:Concurrence_all}
\end{figure}

Knowing that in LO the pair $f \bar f$ is formed in a spin state with maximal entanglement, and hence $C[\rho_{\rm LO}] = 1$, we compute the concurrence in the case of a $t \bar t$ pair in the collinear limit as a function of $\beta$ in Fig.~\ref{fig:Concurrence_all}. This decoherence appears as a reduction in the concurrence.

The scalar emission, which is solely an identity map, does not change the entanglement. For other cases, as expected from NLO spin correlation studies, the decoherence effects are suppressed due to the power $\alpha_\Gamma$. However, they do contribute at a high value of $\beta$. 

These effects are expected to play a role in precision studies of entanglement at colliders. 
One can see that for a QCD-type interaction ($\alpha_{\rm V} \sim 0.1$, blue dashed), a 1\% decrease is expected. The reason behind this effect being small is two-fold: higher-order in perturbation theory and the collinear radiation being suppressed $1/m_t$. However, to assess their real impact on colliders, a complete phenomenology study is required. 

\section{Splitting functions as Kraus operators}
\label{sec:Correspondence}
We now show that the Kraus operators are directly related to the Altarelli-Parisi (AP) splitting functions~\cite{ALTARELLI1977298} for $q\rightarrow qg$. This identification connects a quantum information formalism to a well-known particle physics one.

Following the discussion in~\cite{Richardson:2018pvo,Hamilton:2021dyz}, let us consider a production $n$-point amplitude $\mathcal{M}_n$ which undergoes a splitting $\tilde{\imath}\rightarrow ik$ of particle $\tilde{\imath}$ to an $(n+1)$-point amplitude $\mathcal{M}_{n+1}$. The factorization is given by
\begin{align}
    \mathcal{M}^{\lambda_i\lambda_k}_{n+1}({\cdots}, p_i,p_k,{\cdots}) =\mathcal{S}^{\lambda_{\tilde{\imath}}\lambda_i\lambda_k}_{\tilde{\imath}\rightarrow ik}\,\,\mathcal{M}^{\lambda_{\tilde{\imath}}}_{n}({\cdots},p_{\tilde{\imath}},{\cdots})\,,
\end{align}
where the effective QCD splitting amplitudes $\mathcal{S}^{\lambda_{\tilde{\imath}}\lambda_i\lambda_k}_{\tilde{\imath}\rightarrow ik}$ are given by
\begin{align}
\mathcal{S}^{\lambda_{\tilde{\imath}}\lambda_i\lambda_k}_{\tilde{\imath}\rightarrow ik} = \frac{1}{\sqrt{2}}\frac{g_s}{p_i\cdot p_k} {\cal F}^{\lambda_{\tilde{\imath}}\lambda_i\lambda_k}_{\tilde{\imath}\rightarrow ik} (z) S_\tau(p_i,p_k)\,,
\end{align}
where $z$ is the fraction of particle $i$ momentum relative to $\tilde{\imath}$. The spinor product $S_{\tau}(p_i,p_k)$ depends on the azimuthal angle between $ik$ plane and the spin index $\tau= \pm$. For our purposes, the relevant splitting functions are the $q\rightarrow qg$ ones~\cite{ALTARELLI1977298,Richardson:2018pvo}
\begin{align}
{\cal F}^{\lambda_{\tilde{\imath}},\lambda_i,\lambda_k}_{{\tilde{\imath}}\to i k}
{=} \frac{1}{\sqrt{1-z}} 
\begin{cases}
1 \gamma, & \text{for} \quad(\lambda_{\tilde{\imath}} {=} \lambda_i {=} \lambda_k)\\
z \gamma, & \text{for} \quad(\lambda_{\tilde{\imath}} {=} \lambda_i{=} {-}\lambda_k)\\
\frac{1-z}{z}\frac{m}{q}, &
\text{for} \quad(\lambda_{\tilde{\imath}} {=} {-}\lambda_i{=} \lambda_k)\\
0 & \text{otherwise}
\end{cases} \,, 
\end{align}
where $\gamma = (1-m^2/z^2q^2)^{1/2}$. The massless case can be obtained by simply taking the quark mass $m\to 0$ in the previous equation.
The spin density matrix is obtained by multiplying with the complex conjugate of the amplitude, here simply represented by a bar. The density matrix before the splitting for the momenta of particle $p_{\tilde{\imath}}$ is given by the helicity index ($\lambda_i$ and $\lambda_i'$)
\begin{align}
\rho^{\lambda_{\tilde{\imath}}\lambda_{\tilde{\imath}}'} = \frac{1}{\mathcal{N}_i}
\,\,\mathcal{M}^{\lambda_{\tilde{\imath}}}({\cdots},p_{\tilde{\imath}},{\cdots}) \,\,
\overline{\mathcal{M}}^{\lambda'_{\tilde{\imath}}}({\cdots},p_{\tilde{\imath}},{\cdots})\,.
\end{align}
Similarly, we can define the density matrix after it undergoes a splitting, which now contains and extra helicity index for the outgoing radiation ($\lambda_k$ and $\lambda_k'$)
\begin{align}
&\rho^{(\lambda_{i}\lambda_k)(\lambda_{i}'\lambda_k')} \\
&\quad= \frac{1}{\mathcal{N}_{\tilde{\imath}k}}\mathcal{M}^{\lambda_i\lambda_k}(\cdots, p_i,p_k,\cdots)\,\,\overline{\mathcal{M}}^{\lambda'_i\lambda'_k}(\cdots, p_i,p_k,\cdots)\notag \,. 
\end{align}
In the collinear limit one can write  the following relation between the two density matrices
\begin{align}
\rho^{(\lambda_{i}\lambda_k)(\lambda_{i}'\lambda_k')} = \bigg[\mathcal{S}^{\lambda_{\tilde{\imath}}\lambda_i\lambda_k}_{\tilde{\imath}\rightarrow ik}\bigg]
\,\,
\rho^{\lambda_{\tilde{\imath}}\lambda_{\tilde{\imath}}'}
\,\,
\bigg[\mathcal{S}^{\lambda'_{\tilde{\imath}}\lambda'_i\lambda'_k}_{\tilde{\imath}\rightarrow ik}\bigg]^\dagger 
\left(\frac{\mathcal{N}_{i}}{\mathcal{N}_{\tilde{\imath}k}}\right) \,. 
\end{align}
So far this corresponds to what is usually done when including spin correlations in parton shower evolution~\cite{COLLINS1988794,KNOWLES1988767,Richardson:2018pvo,Hamilton:2021dyz}. In our case, we want to model the scenario where this collinear emission is unresolvable by the detector.
For that, we need to obtain the reduced density matrix and understand the emission as decoherence effects. Tracing over the radiation degrees of freedom, we obtain
\begin{align}
\rho^{\lambda_{i}\lambda_{i}'}_{\rm red} =
\sum_{\sigma,\sigma'=\pm}\delta_{\sigma\sigma'} 
\int_{p_k}
\langle p_k,\sigma|
\rho^{(\lambda_{i}\lambda_k)(\lambda_{i}'\lambda_k')} | p_k,\sigma'\rangle
\,. \end{align}
Here, we use a compact notation $\sigma$ to indicate the sum of the possible channels, which coincides with the helicities (either positive or negative) of the emitted massless radiation. Then, the reduced density matrix can be written as
\begin{align}
\rho^{\lambda_{i}\lambda_{i}'}_{\rm red} =
\sum_{\sigma=\pm}
\int_{p_k}
\mathcal{S}^{\lambda_{\tilde{\imath}}\lambda_i\sigma}_{\tilde{\imath}\rightarrow ik}
\cdot
\rho^{\lambda_{\tilde{\imath}}\lambda_{\tilde{\imath}}'}
\cdot
\mathcal{S}^{\lambda'_{\tilde{\imath}}\lambda'_i\sigma}_{\tilde{\imath}\rightarrow ik} \,,
\end{align}
where $\mathcal{S}^{\lambda_{\tilde{\imath}}\lambda_i\sigma}_{\tilde{\imath}\rightarrow ik}$ can be directly related to the Kraus operators. 
We can now define the collinear emission as a part of quantum map
\begin{align}
\bar{\mathcal{E}}_{\rm col}[\rho] = 
\sum_{\sigma=\pm}
\int_{p_k}
\mathcal{S}^{\lambda_{\tilde{\imath}}\lambda_i\sigma}_{\tilde{\imath}\rightarrow ik}
\cdot 
\rho^{\lambda_{\tilde{\imath}}\lambda_{\tilde{\imath}}'}
\cdot 
\mathcal{S}^{\lambda'_{\tilde{\imath}}\lambda'_i\sigma}_{\tilde{\imath}\rightarrow ik} \,.
\end{align}
Note that the map above acts solely on particle $\tilde{\imath}$, \textit{i.e.}, is ``local'' in the particles, a well-known property of collinear radiation. In order to include it in the evolution of a bipartite system as in the previous section, we just augment it with the identity $\mathds{1}_2$. This structure is the same as in Eq.~\ref{eq:Map_Hard_Real} for the vector case. We finally arrive to the  identification 
\begin{align}
{\sf q}^{\rm hard}\sum_{j\neq \text{id}} &K_{j} \rho_{\rm LO} K^\dagger_{j} = \\
&
\sum_{\sigma=\pm}
\int_{p_k}
\bigg(\mathcal{S}^{\lambda_{\tilde{\imath}}\lambda_i\sigma}_{\tilde{\imath}\rightarrow ik}\otimes \mathds{1}\bigg)
\rho^{\lambda_{\tilde{\imath}}\lambda_{\tilde{\imath}}'}_{\rm LO}
\bigg(\mathds{1}\otimes 
\mathcal{S}^{\lambda'_{\tilde{\imath}}\lambda'_i\sigma}_{\tilde{\imath}\rightarrow ik}\bigg)\,. \nonumber
\end{align}
Note that because of the factorization in the collinear limit, the integral over $p_k$ does not act on the initial density matrix. To generalize this map, one can consider next-to-soft and next-to-collinear emissions which will lead to next-to-leading-log terms. One could also consider emissions from both $f$ and $\bar f$, which are formally higher-order as well as higher loops or double emission. All these effects will then generalize the derived maps and will be explored in forthcoming studies.
\vspace*{.2cm}
\section{Conclusion}
\label{sec:Conclusions}
The study of entanglement at colliders is evolving into a more established research field. Quantum effects that have been so far neglected, such as decoherence due to strong or electroweak radiation and in general higher-order corrections in the gauge couplings need to be considered when comparing theory and data. A formalism that interprets such radiation effects in particle physics in terms of quantum information process is therefore needed. 

In this study, we have demonstrated a proof-of-concept approach to modeling decoherence effects arising from soft and collinear radiation in spin entanglement at colliders, using standard quantum field theory techniques. Employing the decay of a scalar boson to a $f\bar f$ pair as a simplified example, we showed that radiative emissions, both real and virtual, can be represented as quantum maps that introduce decoherence effects. Notably, we identified a correspondence between Kraus operators and the integrated Altarelli-Parisi splitting functions. This framework could prove valuable for quantum accurate simulations of parton showers, see, {\it e.g.},~\cite{Karlberg:2021kwr}. Furthermore, it would be worthwhile to extend this correspondence to higher-order splitting functions.

 While centered on a simplified example, this initial exploration outlines a general approach that can be broadly applied to the study of decoherence in other entangled systems. Using our framework, it is possible to estimate the magnitude of such effects at colliders. The simple case presented here could already be used for \( t\bar{t} \) production at threshold—where the pair is in a spin-singlet scalar state—as well as in \( h \to \tau^+ \tau^- \). Applications to other \( f\bar{f} \) final states, such as \( e^+e^- \rightarrow \tau^+\tau^- \), \( t\bar{t} \), or \( pp \rightarrow \tau^+\tau^- \), \( t\bar{t} \), are straightforward. These, along with the study of higher-order effects, are left for future work.

\vspace*{.2cm}

\begin{acknowledgments}

We thank Giuseppe De Laurentis, Einan Gardi, Franz Herzog, Jordan Wilson-Gerow for useful conversations.
L.S. thanks the Higgs Centre for Theoretical Physics for the hospitality during his visit to Edinburgh in 2024.
R.A. is supported by UK Research and Innovation (UKRI) under the UK government’s
Horizon Europe Marie Sklodowska-Curie funding guarantee grant [EP/Z000947/1]
and by the STFC grant “Particle Theory at the Higgs Centre”.
The work of AJB is funded in part through STFC grants ST/R002444/1 and ST/S000933/1.
F.M. is partially supported by the PRIN2022 Grant 2022RXEZCJ and  by the project QIHEP-Exploring the foundations of quantum information in particle physics - financed through the PNRR, funded by the European Union – NextGenerationEU, in the context of the extended partnership PE00000023 NQSTI - CUP J33C24001210007. Some manipulations were performed with the help of \texttt{FeynCalc}~\cite{Kublbeck:1992mt,Shtabovenko:2016sxi,Shtabovenko:2020gxv}.
\end{acknowledgments}



\bibliography{main}

\clearpage

\onecolumngrid

\appendix*

\begin{center}
  \textbf{\large Supplemental Material for ``Decoherence effects at colliders from radiation"}\\[.2cm]
  \vspace{0.05in}
  {Rafael Aoude, Alan J. Barr,  Fabio Maltoni and Leonardo Satrioni}
\end{center}

\setcounter{equation}{0}
\setcounter{figure}{0}
\setcounter{page}{1}
\makeatletter
\renewcommand{\theequation}{S\arabic{equation}}
In this Supplemental Material, we provide details on the calculation of the main body as the explicit coefficients of the Kraus operators.
\section*{General coefficients for Kraus operators}
Adding the next-to-leading order \textbf{real} and \textbf{virtual} contributions to the $R$-matrix give the following perturbation
\begin{align}
R = R_{\rm LO} + R_{\rm NLO} = 
R = R_{\rm LO} + R_{\rm NLO}^{\rm real} + R_{\rm NLO}^{\rm virt.}
\end{align}
But we want to write as a quantum map acting on the leading density matrix $\rho_{\rm LO}$. 
Given that we choose the initial state to be normalized, $\text{tr}[\rho_{\rm LO}]=1$, the final density matrix is then given by
\begin{align}
\rho = \frac{1}{\text{tr}[R]} R = \sum_j K_j\, \rho_{\rm LO}\, K_j^\dagger 
= {\sf p}_{\rm id.} \mathds{1} \rho_{\rm LO} \mathds{1} + {\sf q} \sum_{j\neq \rm id} K_j \rho_{\rm LO} K^\dagger_j
\end{align}
The virtual correction ends up being an identity operator due to our particular case of a scalar decay. Thus,
\begin{align}
 R_{\rm NLO}^{\rm virt.} = \tilde{\sf p}_{\rm virt.} \mathds{1} R_{\rm LO}\mathds{1}
\qquad\Rightarrow\qquad
\frac{1}{\text{tr}[R]}
 R_{\rm NLO}^{\rm virt.}
=\tilde{\sf p}_{\rm virt.}\frac{\text{tr}[R_{\rm LO}]}{\text{tr}[R]}\mathds{1} \rho_{\rm LO}\mathds{1}
\end{align}
whereas in~\cite{Afik:2020onf}, tilde and non-tilde coefficients are related by a normalization.
The contribution from the virtual radiation is given by
\begin{align}
\tilde{\sf p}_{\rm virt.}^{\rm S} &= \frac{\alpha_{\rm S}}{4\pi}\frac{1}{\beta^2} \left( 8m_t^2B_0(m_t^2,0,m_t^2) -4m_t^2\beta^2C_0(m_t^2,m_t^2,m_\varphi^2,m_t^2,0,m_t^2) - \beta^2 B_0(m_\varphi^2,m_t^2,m_t^2) \right)\\ 
\tilde{\sf p}_{\rm virt.}^{\rm P}  &= \frac{\alpha_{\rm P}}{4\pi} \frac{1}{9}B_0(m_{\varphi}^2,m_t^2,m_t^2)\\
\tilde{\sf p}_{\rm virt.}^{\rm V} &= \frac{2\alpha_{\rm V}}{9\beta^2 m_\varphi^2}\bigg(
2(m_\varphi^2-2m_t^2)(2 B_0(m_t^2,0,m_t^2) - \beta^2m_\varphi^2 C_0(m_t^2,m_t^2,m_\varphi^2,m_t^2,0,m_t^2))
- 8m_t^2 B_0(m_\varphi^2,m_t^2,m_t^2)
\bigg)\\ 
\tilde{\sf p}_{\rm virt.}^{\rm A} &= 
\frac{\alpha_{\rm A}}{4\pi\beta^2m_\varphi^2}
\bigg(
2(m_\varphi^2-6m_t^2)(2 B_0(m_t^2,0,m_t^2) - \beta^2m_\varphi^2 C_0(m_t^2,m_t^2,m_\varphi^2,m_t^2,0,m_t^2))
+ 8m_t^2 B_0(m_\varphi^2,m_t^2,m_t^2)\bigg)
\end{align}
where $B_0$ and $C_0$ are the typical Passarino-Veltman one-loop functions~\cite{Passarino:1978jh}. These functions are then added by the quark mass and the wavefunction renormalization constants to obtain the result free from UV divergences. We have checked that these results match the literature, {\it e.g.},~\cite{Kniehl:1991ze} for the vector case). It is interesting to note that the pseudoscalar is proportional just to $B_0(m_{\varphi}^2,m_t^2,m_t^2)$, hence free of IR divergencies, as noted in~\cite{Maltoni:2024wyh}.

For the real emission, we can split into its soft and hard part by the following split into the emitted radiation momentum integral
\begin{align}
\text{tr}_{\mathcal{H}_k}
[\cdot ]=\int \dd \Phi(p_k)
\sum_{\sigma=\pm} \langle p_k,\sigma|\cdot |p_k,\sigma\rangle =
\int_{E_k\leq \omega_0}\!\! \dd \Phi(p_k)
\sum_{\sigma=\pm} \langle p_k,\sigma|\cdot |p_k,\sigma\rangle
+
\int_{E_k> \omega_0} \!\! \dd \Phi(p_k)
\sum_{\sigma=\pm} \langle p_k,\sigma|\cdot |p_k,\sigma\rangle
\end{align}
given by an energy scale $E_k \leq \omega_0$ for soft radiation. For the soft part, we use leading-order soft theorem, the amplitude factorizes into a soft factor and the tree-level decay amplitude. The former is a scalar function in spin space and therefore does not affect the LO density matrix.
For the integral, there will be a contribution from the "hard" part. We take this integrand to be collinear, which will have a scalar contribution, \text{i.e.}, the identity part of the part, and the interesting non-trivial Kraus operator contribution. Using power counting arguments for the integrand, one can see that the latter one is free of divergencies.

\section*{Explicit Kraus operators}
\label{sec:Explicit_Kraus}

The full structure of the real emission $\mathcal{E}_{\rm R} [\rho_{\rm LO}]$ (soft+hard) for each contribution $g_\Gamma \Gamma^\mu$ is given by the following changes of the density matrix $\Delta \rho$
\begin{align}
\bar{\mathcal{E}}_{\rm R} [\rho_{\rm LO}]\, {=}
\alpha_\Gamma
\begin{pmatrix}
(\Delta\rho_{\Gamma})_{11} &0 &0 &0\\
0 &(\Delta\rho_{\Gamma} )_{22} &(\Delta\rho_{\Gamma} )_{23} &0\\
0 &(\Delta\rho_{\Gamma} )_{23} &(\Delta\rho_{\Gamma} )_{22} &0\\
0 &0 &0 &(\Delta\rho_{\Gamma} )_{11}
\end{pmatrix}
\end{align}
where $\alpha_\Gamma = g_\Gamma^2/4\pi$ and the element $(\Delta\rho_{\Gamma})_{ij}$ is just to represent the non-zero entries. For each case, we have that
\begin{align}
   &\text{P}:\quad (\Delta\rho_{\rm P})_{11} = (\Delta\rho_{\rm P})_{22} = 0 \neq (\Delta\rho_{\rm P})_{23}\\
    &\text{V}:\quad (\Delta\rho_{\rm V})_{11} = (\Delta\rho_{\rm V})_{22} = (\Delta\rho_{\rm V})_{23} \\
    &\text{A}:\quad (\Delta\rho_{\rm A})_{11} = (\Delta\rho_{\rm A})_{22} \neq (\Delta\rho_{\rm P})_{23}
\end{align}
The Kraus operators that give the real emission NLO correction are
\begin{align}
    \text{pseudo}: &\quad K_{03} = \zeta_z \mathds{1}_2\otimes \sigma_3 \\
    \text{vector}: &\quad K_{0+}  = \zeta_+ \mathds{1}_2\otimes \sigma^+ \quad \text{and} \quad K_{0-}  = \zeta_- \mathds{1}_2\otimes \sigma^- \\
        \text{axial}: &\quad K_{0+}  = \zeta_+ \mathds{1}_2\otimes \sigma^+ \quad \text{and} \quad K_{0-}  = \zeta_- \mathds{1}_2\otimes \sigma^- \qquad  \text{and} \quad K_{03} = \zeta_z \mathds{1}_2\otimes \sigma_3 
\end{align}
where this Kraus are acting non-trivially on the $\bar{t}$ emission. Here we have written the Kraus operators with $\sigma_\pm = (\sigma^1 \pm i \sigma^2)$.
Similarly, we have the operators acting on the $t$ where the second index is $0$. Using the usual quantum information language, we can see that the pseudo acts as a phase-flip gate, while the vector acts as a combination from bit-flip and bit-phase-flip.

\end{document}